\documentstyle[floats,prl,aps]{revtex} 
\begin{document}

\renewcommand{\thefootnote}{\fnsymbol{footnote}}
\draft

\title{\large\bf Twisted Parafermions }

\author{Xiang-Mao Ding $^{a,b}$ \footnote{E-mail:xmding@maths.uq.edu.au},
Mark D. Gould $^a$ and Yao-Zhong Zhang $^a$ \footnote{E-mail: 
yzz@maths.uq.edu.au}}          

\address{$^a$ Center of Mathematical Physics, Department of Mathematics, \\
University of Queensland, Brisbane, Qld 4072, Australia.\\
$^b$ Institute of Applied Mathematics, Academy of Mathematics 
and System Sciences; \\
Chinese Academy of Sciences, P.O.Box 2734, 100080, China.}

\maketitle

\vspace{10pt}

\def\beq{\begin{equation}}
\def\eeq{\end{equation}}
\def\beqa{\begin{eqnarray}}
\def\eeqa{\end{eqnarray}}         
\def\beqas{\begin{eqnarray*}}
\def\eeqas{\end{eqnarray*}}
\def\bea{\begin{array}}
\def\eea{\end{array}}   
\def\bds{\begin{displaymath}}
\def\eds{\end{displaymath}}

\def\nn{\nonumber}
\def\no{\noindent}

\def\bebb{\begin{thebibliography}}
\def\eebb{\end{thebibliography}}      
\def\bbit{\bibitem}


\def\dl{\delta}
\def\Dl{\Delta}

\def\eps{\epsilon}
\def\vpsn{\varepsilon}

\def\Gm{\Gamma}
\def\gm{\gamma}

\def\lm{\lambda}
\def\Lm{\Lambda}

\def\sgm{\sigma}
\def\Sgm{\Sigma}
\def\vsgm{\varsigma}




\def\tl{\tilde}  
\def\empst{\emptyset}
\def\p{\partial}
\def\nft{\infty}


\def\ot{\otimes}
\def\op{\oplus}

\def\dg{\dagger}
\def\eqv{\equiv}

\def\psd{\psi^{\dg}}
\def\Psd{\Psi^{\dg}}
\def\journal#1&#2(#3){\unskip, \sl #1\ \bf #2 \rm(#3) }
\def\andjournal#1&#2(#3){\sl #1~\bf #2 \rm (#3) }
\def\npb#1#2#3{Nucl. Phys. {\bf B#1}, #2 (#3)}
\def\pl#1#2#3{Phys. Lett. {\bf #1B}, (#2) #3}
\def\plb#1#2#3{Phys. Lett. {\bf B#1}, (#2) #3}
\def\prl#1#2#3{Phys. Rev. Lett. {\bf #1}, (#2) #3}
\def\physrev#1#2#3{Phys. Rev. {\bf D#1}, (#2) #3}
\def\prd#1#2#3{Phys. Rev. {\bf D#1} (#2) #3}
\def\ap#1#2#3{Ann. Phys. {\bf #1}, (#2) #3}
\def\prep#1#2#3{Phys. Rep. {\bf #1}, (#2) #3}
\def\rmp#1#2#3{Rev. Mod. Phys. {\bf #1}, (#2) #3}
\def\cmp#1#2#3{Commun. Math. Phys. {\bf #1}, (#2) #3}
\def\cqg#1#2#3{Class. Quant. Grav. {\bf #1}, (#2) #3}
\def\mpl#1#2#3{Mod. Phys. Lett. {\bf A#1}, (#2) #3}
\def\lmp#1#2#3{Lett. Math. Phys. {\bf #1}, (#2) #3}

\begin{abstract}
A new type of nonlocal currents (quasi-particles), which we call
twisted parafermions, and its corresponding twisted $Z$-algebra 
are found. The system consists of one spin-1 bosonic field and 
six nonlocal fields of fractional spins. Jacobi-type identities 
for the twisted parafermions are derived, and a new conformal 
field theory is constructed from these currents. As an application, 
a parafermionic representation of the twisted affine current algebra 
$A^{(2)}_2$ is given.
\end{abstract}

\pacs{11.25.Hf; 03.65.Fd; 11.10.Lm.} 


The introduction of the $Z_k$ parafermions \cite{ZaFa} in the context of 
statistical models and conformal field theory \cite{BPZ}
is perhaps one of the most significant
conceptual advances in modern theoretical physics. From field theory
point of view, parafermions generalize the Majorano fermions and have
found important applications in superstring 
theory~\cite{ZQiu,FrWe}, fractional supersymmetry and fractional 
superstring~\cite{Tye}, $T$-duality~\cite{AAB} and mirror symmetry~
\cite{Yau}.  In a very recent work 
by Maldacena, Moore and Seiberg, $D$-branes were constructed with the 
help of the $Z_k$ parafermions~\cite{MMS}. From statistical physics
point of view, parafermions  are related to the
exclusion statistics introduced by Haldane~\cite{Hald}. In particular, 
the $Z_k$ parafermion models offer various extensions
of the Ising model which corresponds to the $k=2$ case~\cite{ZaFa}. 
Other examples include the $3$-state Potts model ($k=3$)~\cite{ZaFa,FaZa} 
and the Ashkin-Teller model ($k=4$)~\cite{Yang}. Parafermions also have
applications in Quantum Hall Effects~\cite{Cap}, Bose-Einstein 
Condensates~\cite{Coop} and Quantum Computations~\cite{Wu}.

The category for parafermions (nonlocal operators) is the generalized 
vertex operator algebra \cite{DL,LW}. 
The $Z_k$ parafermion algebra was referred 
to as $Z$-algebra in \cite{DL,LW}, and the $Z_k$ parafermions are 
canonically modified $Z$-algebras acting on certain quotient spaces 
$A_1 ^{(1)}$-modules 
defined by the action of an infinite cyclic group. It was proved that the 
$Z$-algebra is identical with the $A^{(1)}_1$ parafermion. The $Z_k$ 
parafermion characters and their singular vectors were studied in~\cite{JM}. 

The imoprtance of parafermions inspired many researchers to study
various extensions of the $Z_k$ parafermions which are basically related
to the simplest $A_1^{(1)}$ algebra. Gepner proposed a
parafermion algebra associated with any given untwisted 
affine Lie algebra ${\cal G}^{(1)}$~\cite{Gepn,Gepn2}, which has been
subsequently used in the study of $D$-branes. The operator product
expansions (OPEs) and the corresponding $Z$-algebra of the untwisted
parafermions were studied in \cite{DFSW,WaDi}.

In this paper, we find a new type of nonlocal currents (quasi-particles), 
which will be referred to as twisted parafermions.
The system contains a bosonic spin-$1$ field and six nonlocal fields with 
fractional spins. Some of the fields are in the Ramond sector and 
some are in the Neveu-Schwarz (NS) sector. They correspond to
a new type of qusi-particles or generalized Majorana fermions. We derive 
the corresponding twisted $Z$-algebra and the Jacobi-type identities for 
the twisted parafermion currents. From the twisted parafermions, we construct 
a new conformal field theory which is different from the known ones. 
As an application, we obtain a parafermionic representation of the 
twisted affine current algebra $A^{(2)}_2$, which we expect to have 
application in the description of the entropy of the $Ads_3$ balck hole
\cite{FeMa}.


It is well-known that Euclidian correlation functions are defined only 
if operators in the correlators are time-ordered~\cite{ItZ2}. In the radial 
picture, $|z|>|w|$ means that $z$ is later than $w$. 
In the Euclidian functional integral definition of correlation functions, 
the time ordering is automatic. So, in Euclidian field theory the operator 
products $A(z)B(w)$ are only defined for $|z|>|w|$. Therefore the radial 
ordering is implied throughout this paper. 

Now we propose the twisted parafermion current algebra: 

\beqa
\psi _l(z)\psi _{l^{\prime}}(w)(z-w)^{ll^{\prime} /2k}
&&=\frac{\delta _{l+l^{\prime},0}}{(z-w)^2}
+\frac{\varepsilon _{l,l^{\prime}}}{z-w}
\psi _{l+l^{\prime}}(w)+\cdots, \nn\\
\psi _{\tl l}(z)\psi _{\tl {l^{\prime}}}(w)(z-w)^{ll^{\prime} /2k}
&&=\frac{\delta _{\tl{l}+\tl{l^{\prime}},0}}{(z-w)^2}
+\frac{\varepsilon _{\tl{l},\tl{l^{\prime}}}}{z-w}
\psi _{l+l^{\prime}}(w)+\cdots, \\
\label{eq:twst}
\psi _l(z)\psi _{\tl{l^{\prime}}}(w)(z-w)^{ll^{\prime} /2k}
&&=\frac{\varepsilon _{l,l^{\prime}}}{z-w}
\psi _{\tl{l}+\tl{l^{\prime}}}(w)+\cdots,\nn
\eeqa

\no where $l,\;l^{\prime}=\pm 1$ and $\tl{l}$, $\tl{l^\prime}=
\tl{0},\pm\tl{1},\pm\tl{2}$; $\varepsilon _{l,l^{\prime}},~
\varepsilon _{\tl{l},\tl{l^{\prime}}}$ and $\varepsilon _{l,\tl{l^{\prime}}}$ 
are structure constants. 
If we denote $\psi _l$ or $\psi _{\tl{l}}$ by $\Psi _a$, 
then we can rewrite the above relations as:

\beq
\Psi _a(z)\Psi _b(w) (z-w)^{ab/2k}
\equiv \sum _{n=-2} ^{\infty}(z-w)^n [\Psi _a\Psi _b]_{-n},
\label{eq:Par}
\eeq
  
\no where $a,b=\tl{0},\pm 1,\pm\tl{1},\pm\tl{2}$. So we have $
 \left[ \Psi _{a}\Psi _{b} \right] _l=0\;(l\geq 3)$, $
 \left[ \Psi _{a}\Psi _{b} \right]_2=\delta _{a+b,0}$ and $
 \left[\Psi _{a}\Psi _{b} \right]_1={\varepsilon _{a,b}}\Psi _{a +b}$. 
For consistency, $\varepsilon _{a,b}$ must have the properties: 
$\varepsilon _{a,b}=
-\varepsilon _{b,a}=-\varepsilon _{-a,-b}
=\varepsilon _{-a,a+b}$ and $\varepsilon _{a,-a}=0 $. Due to the mutually 
semilocal property between two parafermions, the radial ordering products 
are multivalued functions. So we define the radial ordering product of 
(generating) twisted parafermions(TPFs) as

\beq
\Psi _{a}(z)\Psi _{b}(w)(z-w)^{ab/2k}
=\Psi _{b}(w)\Psi _{a}(z)(w-z)^{ab/2k},
\label{eq:rlrr}
\eeq

\noindent which, like the untwisted case, is an extension of the fermion 
(i.e. $ab=1,\;k=1$) and boson (i.e. $k\rightarrow \infty $) theories. 

For every field in the parafermion theory there are a pair of charges 
$(\lambda, \bar{\lambda})$, which take values in the weight lattice. 
We denote such a field by $\phi _{\lambda, \bar{\lambda}}(z,\bar{z})$ 
\cite{ZaFa,Gepn,DFSW}. The OPE of $\Psi _{a}$ with 
$\phi _{\lambda, \bar{\lambda}}(z,\bar{z})$ is given by 
 
\beq
\Psi _{a}(z)\phi _{\lambda, \bar{\lambda}}(w,\bar{w})
=\sum _{m=-\infty} ^{\infty}(z-w)^{-m-1-a\lambda/2k}
A_m ^{a,\lambda}
\phi _{\lambda, \bar{\lambda}}(w,\bar{w}),
\eeq

\noindent where $m \in Z$ (Ramond sector) for $a=l$ and 
$m \in Z+\frac{1}{2}$ (Neveu-Schwarz sector) for $a=\tl{l}$. The 
action of $A_m ^{a,\lambda}$ on $\phi _{\lambda, \bar{\lambda}}(z)$ 
is defined by the integration

\beq
A_m ^{a,\lambda}\phi _{\lambda, \bar{\lambda}}(w,\bar{w})
=\oint _{c_w}\;dz\;(z-w)^{m+a\lambda/2k}
\Psi _{a}(z)\phi _{\lambda, \bar{\lambda}}(w,\bar{w}),
\eeq

\noindent where $c_w$ is a contour around $w$. 
Consider the difference of integrals  

\beqa
& &\oint\oint\;du\;dz \Psi _{a}(u) \Psi _{b}(z)
\phi _{\lambda, \bar{\lambda}}(w,\bar{w})(u-z)^{-p-1+ab/2k} \nn\\
&&~~~\times(u-w)^{m+q+1+a\lambda/2k}(z-w)^{n+p-q+b\lambda/2k},
\eeqa

\noindent along two contours satisfying $|u-w|>|z-w|$ and $|u-w|<|z-w|$, 
respectively. The difference of the two contour integrals can be expressed 
as a two-fold integration of a $u$-contour enclosing $z$ once followed by a
$z$-contour enclosing $w$ once. Properly carrying out the Taylor expansion 
of $(u-z)^x$, we then obtain the so-called twisted $Z$-algebra relations,      

\beqa
&&\sum _{l=0} ^{\infty}C_{-p-1+ab/2k} ^{(l)}
 \left[A_{m-l-p+q} ^{a,\lambda+b}A_{n+l+p-q} ^{b,\lambda}
+(-1)^p A_{n-l-q-1} ^{b,\lambda+a}A_{m+l+q+1} ^{a,\lambda}\right]
\nonumber\\
&&=C_{m+q+1+a\lambda/2k} ^{(p+2)}\delta _{a,-b}\delta _{m,-n}
+ C_{m+q+1+ab/2k} ^{(p+1)}\varepsilon _{a,b}
A^{a+b,\lambda} _{m+n}\nn\\
&&~~~+\sum _{r=0} ^{\infty}C^{p-r} _{m+q+1+ab/2k}
 [\Psi _{a}\Psi _{b}]_{-r,m+n} ^{\lambda},
\eeqa

\noindent where $p=2q$ or $2q+1$ and $[\Psi _{a}\Psi _{b}]_{-m,n} ^{\lambda}$ 
is defined by 
\beq
[\Psi _{a}\Psi _{b}]_{-m,n} ^{\lambda}
\phi _{\lambda, \bar{\lambda}}(w,\bar{w})
=\oint _{c_w}\;dz\;(z-w)^{m+n+1+(a+b)\lambda/2k}
 [\Psi _{a}\Psi _{b}]_{-m}(z)
\phi _{\lambda, \bar{\lambda}}(w,\bar{w}). 
\eeq


Let $A_{a}$ and $B_{b}$ be two arbitrary functions of the twisted 
parafermions with charges $a$ and $b$, respectively. These fields are 
local ($a$ or $b=0$) or semilocal ($a$ and $b$ $\not= 0$). The OPEs 
can be written as 

\beq
A_{a}(z)B_{b}(w)(z-w)^{ab/2k}=
\sum_{n=- [h_A+h_B ] }^{\infty} [A_{a}B_{b}]_{-n}(w)(z-w)^n,
\eeq  

\noindent where $ [ h_A ] $ stands for the integral part of the 
dimension of $A$. Hence we have 
$[ A_{a}B_{b}]_n(w)=\oint_w \ dz \ A_{a}(z)B_{b}(w)(z-w)^{n-1+ab/2k}$. 
It is easy to find the following relation between the 
three-fold radial ordering products

\beqa
&&\left\{\oint_w \ du \oint_w \ dz \ R(A(u)R(B(z)C(w)))\right.\nn\\
&&-\oint_w \ dz \oint_w \ du \ (-)^{ab/2k}R(B(z)R(A(u)C(w)))\nonumber\\
&&-\left. \oint_w \ dz \oint_z \ du \ R(R(A(u)B(z))C(w))\right\} \nn\\
&&(z-w)^{p-1+bc/2k}(u-w)^{q-1+ac/2k}(u-z)^{r-1+ab/2k}=0,  
\label{eq:threef}
\eeqa

\noindent where integers $p,\;q,\;r$ are in the regions: $-\infty< \,p\, 
\leq  [h_B+h_C ] ,\;\;-\infty<\,q\,\leq \, [h_C+h_A ] ,\;\;-\infty<\,r\,
\leq\, [h_A+h_B ] $; and $a,\;b,\;c$ are parafermionic charges of the 
fields $A,\;B$ and $C$, respectively. The above equation is an extension 
of $A(BC)-B(AC)-[A,B]C=0$. 
Performing the binomial expansion, we obtain the following twisted 
Jacobi-like identities: 

\beqa
&&\sum_{i=p}^{ [h_B+h_C ] }C_{r-1+ab/2k}^{(i-p)}
[A [BC ]_i ]_{Q-i}(w)+(-)^r\sum_{j=q}^{ [h_C+h_A ] }C_{r-1+ab/2k}^{(j-q)}
[B [AC ]_j ]_{Q-j}(w)\nonumber\\
&&=\sum_{l=r}^{ [h_B+h_A ] }(-)^{(l-r)}C_{q-1+ac/2k}^{(l-r)}
[ [AB ]_lC ]_{Q-l}(w),
\label{eq:Jacobi}
\eeqa

\noindent where $ Q=p+q+r-1,C_x^{(l)}=\frac{(-)^lx(x-1)...(x-l+1)}{l!}$ 
and $C_0^{(0)}=C_n^{(0)}=C_{-1}^{(l)}=1,\;C_p^{(l)}=0$, for $l>p>0$. 
This identity will be used extensively for our purpose.
Performing analytic continuation one obtains 

\beq
 [ BA ]_r(w)=\sum_{t=r}^{ [h_A+h_B ] }\frac{(-)^t}{(t-r)!}
\partial^{t-r} [AB ]_t(w).
\label{eq:analytic}
\eeq

We remark that $A,\;B,\;C$ can be any composite operators and we can 
calculate any coefficient in the OPEs from the fundamental equation 
(\ref{eq:Par}).


For the twisted parafermion theory to be a conformal field theory, 
we require that the spin-2 terms in the OPEs contain a 
energy-momentum tensor. It is obvious that the spin-2 terms in the 
OPE are $\left[\Psi _a \Psi _b\right]_0$. Since the parafermionic 
charge for the energy-momentum tensor should be zero, so the relevant 
terms are $\left[\Psi _a \Psi _{-a}\right]_0$. Note that 
$\left[\Psi _a \Psi _{-a}\right]_0(z)=\left[\Psi _{-a} \Psi _a\right]_0(z)$, 
we calculate the OPE of $[\Psi _{a}\Psi_{-a}]_0$ with $\Psi _{a}$ and 
$[\Psi _{b}\Psi_{-b}]_0$. Setting $Q=p=2,\;q=1$ and $r=0$ in 
(\ref{eq:Jacobi}), we have 

\beq
\left[[\Psi _{a}\Psi_{-a}]_0 \Psi _{b}\right]_2
=\delta _{a,b}\Psi _{a}
+\vpsn _{a, b}\vpsn _{-a,a+b}\Psi _{b}
+(1+{a}^2/2k)\delta _{a,-b}\Psi _{-a}+\frac{ab}{4k}(1-\frac{ab}{2k})\Psi _{b}.
\eeq

\no From the general theory of conformal field theory~\cite{BPZ}, the 
conformal dimension of $\Psi _a$ should be $1-\frac{a^2}{4k}$. So we 
impose the constraints: 

\beq
\sum _a \vpsn _{a, b}\vpsn _{-a,a+b}=\frac{6-b^2}{k},\hskip 0.5cm
\sum _a ab=0, \hskip 0.5cm \sum _a (ab)^2=12b^2.
\eeq

\no One solution to these constraints is given by 
$\varepsilon _{\tl{1},-\tl{2}}=\varepsilon _{1,\tl{1}}
=\varepsilon _{-1,\tl{2}}=\frac{1}{\sqrt{k}}$ and 
$\varepsilon _{1,-\tl{1}}=\varepsilon _{\tl{0},\tl{1}}=\sqrt{\frac{3}{2k}}$. 
Then we have $\sum_a \left[[\Psi _{a}\Psi_{-a}]_0 \Psi _{b}\right]_2
=\frac{2k+6}{k}\left(1-\frac{b^2}{4k}\right)\Psi _b $. Choose a proper 
normalization and write 
\beq
T_{\psi}=\frac{k}{2k+6}\sum _a [\Psi _{a}\Psi_{-a}]_0.
\eeq
Then $[T_{\psi}\Psi _{b}]_2=\left(1-\frac{b ^2}{2k}\right)\Psi _{b}$. 
Repeating the above process, we obtain  

\beqa
\left[[\Psi _{a}\Psi_{-a}]_0 \Psi _{b}\right]_1
&=&\frac{1}{2}\vpsn _{-a, b}\vpsn _{a,-a+b}\p \Psi _{b}
+\frac{1}{2}\vpsn _{a, b}\vpsn _{-a,a+b}\p \Psi _{b} \nn\\
&&+(1+a^2/2k)\delta _{a,b} \p\Psi _{a}
+(1+a^2/2k)\delta _{-a,b} \p\Psi _{-a},
\eeqa

\noindent or equivalently $[T_{\psi}\Psi _{b}]_1=\partial \Psi _{b} $. 
These results can be written in the form of OPEs

\beq
 T_{\psi}(z)\Psi _{b}(w)=\frac{1-b^2/4k}{(z-w)^2}\Psi _{b}(w)
+\frac{1}{z-w}\partial \Psi _{b}(w)+\ldots.
\eeq

\noindent It follows that the conformal dimension of the twisted 
parafermion is $1,\;1-\frac{1}{4k}$ or $1-\frac{1}{k}$, for a given 
level $k$. Carrying out a similar program for $T$, we obtain the OPE: 

\beq
  T_{\psi}(z)T_{\psi}(w)=
     \frac{c_{\rm tpf}/2}{(z-w)^4}+\frac{2T_{\psi}(w)}{(z-w)^2}
      +\frac{\partial T_{\psi}(w)}{z-w}+\ldots,
\eeq

\no where $c_{\rm tpf}=7-\frac{24}{k+3}=\frac{8k}{k+3}-1$ is the central 
charge of the twisted parafermion theory. 

One of the applications of the twisted parafermionic currents is that they 
give a representation of the twisted affine current algebra $A^{(2)}_2$. 
Introduce eight currents:  

\beqa
&&j^+(z)=2{\sqrt k} \psi _1 (z) 
e^{\frac{i}{\sqrt{2k}}\phi _0(z)}, 
\hskip 0.6cm
j^-(z)=2{\sqrt k} \psi _{-1} (z) 
e^{-\frac{i}{\sqrt{2k}}\phi _0(z)}, \nn\\
&&j^0(z)=2{\sqrt {2k}} i\p \phi _0 (z), 
\hskip 1.0cm
J^+(z)=2{\sqrt k} \psi _{\tl{1}} (z) 
e^{\frac{i}{\sqrt{2k}}\phi _0(z)}, \nn\\
&&J^-(z)=2{\sqrt k} \psi _{-\tl{1}} (z) 
e^{-\frac{i}{\sqrt{2k}}\phi _0(z)}, 
\hskip 0.6cm
J^{++}(z)=2{\sqrt k} \psi _{\tl{2}}(z) 
e^{i {\sqrt\frac{2}{k}}\phi _0(z)},\nn\\
&&J^{--}(z)=2{\sqrt k} \psi _{-\tl{2}} (z) 
e^{-i {\sqrt\frac{2}{k}}\phi _0(z)},
\hskip 0.6cm 
J^{0}(z)=2{\sqrt {6k}} \psi _{\tl{0}} (z).\nn 
\eeqa

\no where $\phi _0$ is an $U(1)$ current obeying $\phi _0 (z)
   \phi _0 (w)=-ln (z-w)$. 
Then it can be checked that the above currents satisfy the OPEs of the 
twisted affine currents algebra $A^{(2)}_2$~\cite{DGZ}. 

In summary, we have found a new type of nonlocal currents
(quasi-particles) and corresponding twisted $Z$-algebra. We derive
the Jacobi-type identities for the twisted parafemion currents.
These Jacobi-type identities are expected to be useful in proving some
of the Ramanujan identites, which play an important role in statistical
physics. From the twisted parafermions, we construct a new conformal
field theory, and give a parafermionic representation of twisted current
algebra $A^{(2)}_2$. This representation is expected to be useful
in the description of entropy of the $Ads_3$ black hole.
\\

This work is financially supported by Australian Research Council. 
One of the authors (Ding) is also supported partly by the 
Natural Science Foundation of China and a grant from the AMSS.
\vskip.1in
\noindent{\em Note Added:} P. Mathieu pointed to us the reference~\cite{CRS},
where graded parafermions associated with the $osp(1|2)^{(1)}$ algebra were
studied and fields with conformal dimensions of $1-\frac{1}{4k}$ and 
$1-\frac{1}{k}$ also appeared. However, our twisted parafermion algebra
is quite different from the graded parafermion algebra in \cite{CRS}.
Our theory contains more fields, is unitary and has a different central 
charge.

\bebb{99}

\bibitem{ZaFa}
A.B.Zamolodchikov and V.A.Fateev, Sov. Phys, JETP {\bf 62}, 215 (1985).

\bibitem{BPZ}
A.A. Belavin, A.M. Polyakov, A.B. Zamolodchikov, {\em Nucl. Phys.} 
{\bf B241}, 333(1984).

\bibitem{ZQiu}
Z.Qiu. Phys. Lett.{\bf B 198}, 497 (1987).

\bibitem{FrWe}
M.Freeman and P.West, Phys. Lett.{\bf B314}, 320 (1993); 
P.H. Frampton and J.T. Liu, \prl{70}{130}{1993}.

\bbit{Tye}
P.C. Argyres and S.-H.H. Tye, \prl {67}{3339}{1991}; 
K.R. Dienes, \npb{413}{105}{1994}

\bbit{AAB}
E. Alvarez, L. Alvarez-Games and I. Bakas, {\em Nucl Phys. Proc. Suppl.}
{\bf 46}, 16 (1996).

\bbit{Yau}
{\it Essays on mirror manifolds }, ed. by S.T. Yau, Hong Kong, 
International Press, 1992.

\bbit{MMS}
J. Maldacena, G. Moore and N. Seiberg, {\it JHEP}{\bf 0107}, 046 (2001). 
hep-th/0105038.

\bbit{Hald}
F.D.M. Haldane, \prl {67}{937}{1991}.

\bibitem{FaZa}
V.A.Fateev and A.B.Zamolodchikov, Nucl.Phys.{\bf B280}, 644 (1987).

\bibitem{Yang}
S.K. Yang, Nucl. Phys.{\bf B285}, 639 (1987).

\bbit{Cap}
A. Cappelli, L.S. Georgiev and I.T. Todorov, \npb{599}{499}{2001}.

\bbit{Coop}
N.R. Cooper, N.K. Wilkin and J.M.F. Gunn, \prl{87}{120}{2001}

\bbit{Wu}
L.-A. Wu and D.A. Lidar, {\it QUbits as Parafermion}, quant-ph/0109078.

\bbit{DL}
C.Y. Dong and J. Lepowsky, {\it Generalized Vertex Algebras and 
Relative Vertex Operators}, 
Birkhauser, 1993. 

\bbit{LW}
J. Lepowsky and R. Wilson, {\em Comm. Math. Phys.} {\bf 62 }, (1978), 43-53;  
in: {\em Vertex Operators in Mathematics and Physics, 
Proc. 1983 M.S.R.I. Conference}, ed. by J. Lepowsky et. al, 
Springer-Verlag, New York, 1985, 97-142.

\bbit{JM}
P. Jacob and P. Mathieu, \npb{587}{514}{2000}; hep-th/0006233.

\bibitem{Gepn}
D. Gepner, Nucl. Phys. {\bf B290}, 10 (1987).

\bibitem{Gepn2}
D. Gepner, Phys. Lett. {\bf B199}, 380 (1987).

%
%
%

\bibitem{DFSW}
X.M. Ding, H. Fan, K.J. Shi, P. Wang and C.Y. Zhu, Phys. Rev. Lett. 
{\bf 70}, 2228 (1993); Nucl. Phys.{\bf B422}, 307 (1994). 

\bibitem{WaDi}
P. Wang and X.M. Ding, Phys. Lett.{\bf B335}, 56 (1994). 

\bibitem{FeMa} S. Fernando and F. Mansouri, Phys. Lett. {\bf B505},
   206 (2001).

\bibitem{ItZ2}
C. Itzykson and J.-B. Zuber,  {\it Quantum Field Theory}, 
McGraw-Hill Inc. 1980.

\bbit{DGZ}
X.M. Ding, M.D. Gould and Y.Z. Zhang, Phys. Lett.  {\bf B523}, 
   367 (2001); hep-th/0109009. 

\bbit{CRS}
J. M. Camino, A. V. Ramaollo and J. M. Sanchez de Santos, 
\npb{530}{715}{1998}; hep-th/9805160.

\eebb

\end{document}